\documentclass[reprint,superscriptaddress,showpacs,amsmath, amssymb,aps,pra]{revtex4-1}
\usepackage{graphicx}
\usepackage{dcolumn}
\usepackage{bm}
\usepackage{subfigure}
\usepackage{algorithm}
\usepackage{algpseudocode}
\usepackage{appendix}
\usepackage{float}
\usepackage[braket, qm]{qcircuit}
\usepackage{mathrsfs}
\usepackage{longtable}
\usepackage{amsfonts}
\usepackage{amstext}
\usepackage[usenames]{xcolor}
\usepackage{epsfig}
\usepackage{epstopdf}
\usepackage{mathrsfs}
\usepackage{bm}
\usepackage{amsmath}
\usepackage{amssymb,amsthm}
\usepackage{tikz}
\usepackage{hyperref}

\usepackage{comment}

\hypersetup{colorlinks=true, citecolor=blue, urlcolor=blue, linkcolor=blue}
\begin{document}
\newtheorem{theorem}{Theorem}
\newtheorem{thm}{Theorem}
\newtheorem{cor}[thm]{Corollary}
\newtheorem{lem}[thm]{Lemma}
\newtheorem{prop}[thm]{Proposition}
\newtheorem{propp}{Proposition}
\newtheorem*{pf}{Proof}
\newtheorem{definition}{Definition}
\newtheorem{ex}{Example}
\newtheorem{remark}{Remark}
\newtheorem*{thmm}{Theorem}
\newtheorem*{corr}{Corollary}

\title{Coherence Fraction in Grover Search Algorithm}

\author{Si-Qi Zhou}
\thanks{These authors contributed equally}
\affiliation{School of Mathematical Sciences, MOE-LSC, Shanghai Jiao Tong University, Shanghai, 200240, China}
\affiliation{Shanghai Seres Information Technology Co., Ltd, Shanghai, 200040, China} \affiliation{Shenzhen Institute for Quantum Science and Engineering, Southern University of Science and Technology, Shenzhen, 518055, China}
\author{Hai Jin}
\thanks{These authors contributed equally}
\affiliation{School of Mathematical Sciences, MOE-LSC, Shanghai Jiao Tong University, Shanghai, 200240, China}
\author{Jin-Min Liang}
\thanks{These authors contributed equally}
\affiliation{State Key Laboratory for Mesoscopic Physics, School of Physics, Frontiers Science Center for Nano-optoelectronics, Peking University, Beijing 100871, China}
\author{Shao-Ming Fei}
\email{feishm@cnu.edu.cn}
\affiliation{School of Mathematical Sciences, Capital Normal
University, Beijing 100048, China}
\affiliation{Max Planck Institute for Mathematics in the Sciences - 04103 Leipzig, Germany}
\author{Yunlong Xiao}
\email{xiao\_yunlong@ihpc.a-star.edu.sg}
\affiliation{Quantum Innovation Centre (Q.InC), Agency for Science Technology and Research (A*STAR), 2 Fusionopolis Way, Innovis \#08-03, Singapore 138634, Republic of Singapore}
\affiliation{Institute of High Performance Computing (IHPC), Agency for Science, Technology and Research (A*STAR), 1 Fusionopolis Way, \#16-16 Connexis, Singapore 138632, Republic of Singapore}
\author{Zhihao Ma}
\email{mazhihao@sjtu.edu.cn}
\affiliation{School of Mathematical Sciences, MOE-LSC, Shanghai Jiao Tong University, Shanghai, 200240, China}
\affiliation{Shanghai Seres Information Technology Co., Ltd, Shanghai, 200040, China}
\affiliation{Shenzhen Institute for Quantum Science and Engineering, Southern University of Science and Technology, Shenzhen, 518055, China}

\begin{abstract}
The question of which resources drive the advantages in quantum algorithms has long been a fundamental challenge. While entanglement and coherence are critical to many quantum algorithms, our results indicate that they do not fully explain the quantum advantage achieved by the Grover search algorithm. By introducing a generalized Grover search algorithm, we demonstrate that the success probability depends not only on the querying number of oracles but also on the coherence fraction, which quantifies the fidelity between an arbitrary initial quantum state and the equal superposition state. Additionally, we explore the role of the coherence fraction in the quantum minimization algorithm, which offers a framework for solving complex problems in quantum machine learning. These findings offer insights into the origins of quantum advantage and open pathways for the development of new quantum algorithms.
\end{abstract}
\maketitle
\section{Introduction}
Quantum computers, grounded in the principles of quantum mechanics, have attracted great attention over the past few decades. These devices are theoretically more efficient than the most advanced classical ones for certain problems~\cite{bennett2000quantum}, such as large integer factoring~\cite{365700}, unstructured search~\cite{grover1996fast, PhysRevLett.79.325}, linear equation solving~\cite{PhysRevLett.103.150502} and quantum system simulation~\cite{feynman2018simulating, RevModPhys.86.153}. Leveraging these quantum advantages, quantum computers hold promise for various applications, from advancing artificial intelligence~\cite{biamonte2017quantum, west2023towards} to developing next-generation materials and pharmaceuticals~\cite{8585034, robert2021resource}. Additionally, the quest for quantum computing has inspired the development of new efficient classical algorithms~\cite{tang2019quantum}. While several small-scale proof-of-concept quantum computers have been realized~\cite{ladd2010quantum}, ongoing research is focused on building larger quantum computers with the potential to surpass the classical ones in specific tasks~\cite{harrow2017quantum, neill2018blueprint, yung2019quantum}.

\begin{figure}[h]
\centering
\includegraphics[width=3.4in]{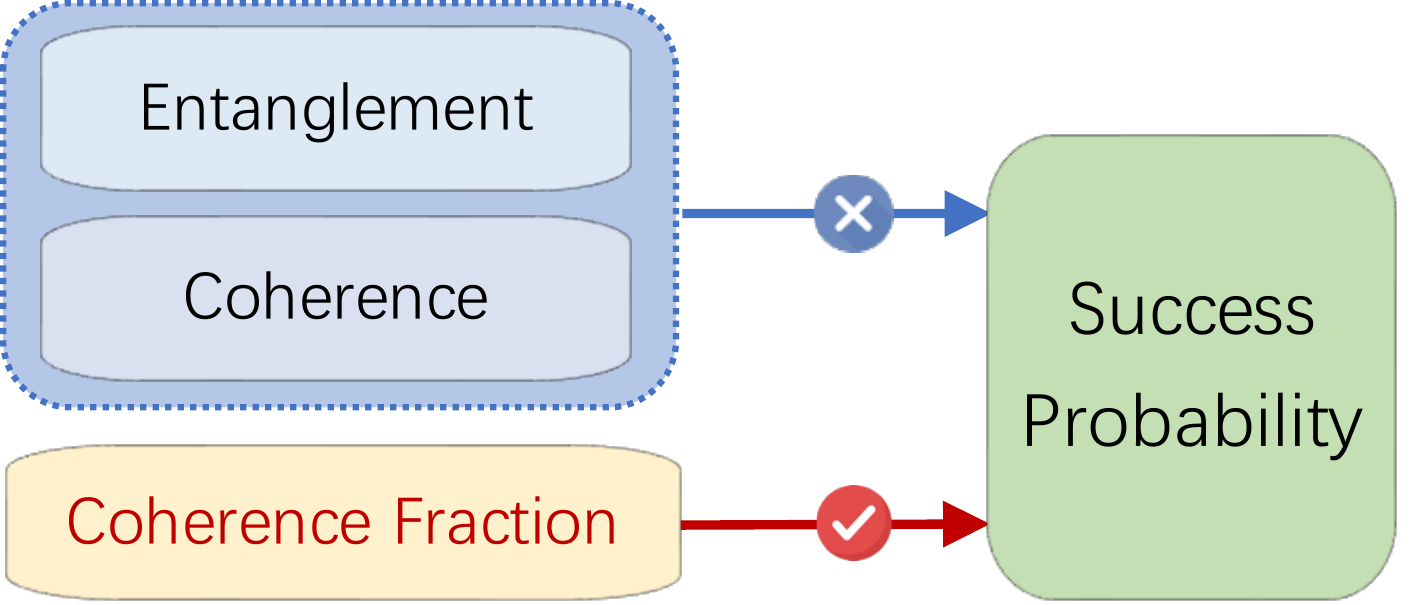}
\caption{\textbf{Diagrammatic sketch of coherence fraction in Grover search algorithm.} The upper bound of the average success probability for the generalized Grover search algorithm depends solely on the coherence fraction, rather than quantum entanglement or coherence.}
\label{fig:CF}
\end{figure}

Quantum computation devices have their advantages in addressing specialized problems \cite{arute2019quantum, zhong2020quantum}. One of the key challenges is investigating which inherent quantum properties of quantum states contribute to the benefits of quantum computation. It has been shown that entanglement and coherence play crucial roles in quantum algorithms \cite{PhysRevA.95.032307, jozsa1998quantum, ekert1998quantum,PhysRevA.100.012349, jozsa2003role, ahnefeld2022coherence, pan2022complementarity}.
Our findings reveal that the average success probability of the generalized Grover search algorithm does not depend on the entanglement or coherence of the initial state. Instead, it depends on the number of oracle queries and the coherence fraction of the initial state, defined by the fidelity between the initial state and the equal superposition state (i.e., maximally coherent state). Once the optimal number of oracle queries is reached, the optimal average success probability depends solely on the coherence fraction of the initial state. A sketch diagram is provided in FIG.~\ref{fig:CF}. In quantum teleportation, there have been similar results regarding the fully entangled fraction (FEF) which measures the optimal fidelity of quantum teleportation~\cite{PhysRevA.54.3824, PhysRevA.54.1838, PhysRevA.60.1888, PhysRevA.62.012311, PhysRevA.66.012301, PhysRevA.91.012310}.

We focus on the specific quantum algorithm with a fixed structure: the Grover search algorithm (\texttt{GSA})~\cite{grover1996fast, PhysRevLett.79.325}. While there have been several proposed generalizations of the \texttt{GSA}~\cite{PhysRevA.64.022307, PhysRevA.68.022326, PhysRevA.71.042320, li2023deterministic, zhou2024distributedexactgeneralizedgrovers}, we revisit the \texttt{GSA} from the source of its quantum advantage. In this work, we consider a generalized version of the Grover search algorithm (\texttt{GGSA}), where the Hadamard gate before the Grover operator is replaced by an arbitrary unitary gate. We provide an exact formulation for the average success probability of \texttt{GGSA}. Our analysis reveals that the upper bound of the average success probability depends solely on the coherence fraction of the initial state. We define the coherence fraction of a state $\rho$ as the Uhlmann's fidelity $F$ between $\rho$ and $|\eta\rangle $,
\begin{align}
    f_{c}(\rho):= F(|\eta\rangle, \rho)=\langle \eta | \rho |\eta\rangle,
\end{align}
where $ |\eta\rangle $ is the equal superposition state or the maximally coherent state. The coherence fraction quantifies how close a quantum state is to a maximally coherent state~\cite{PhysRevA.100.032324, karmakar2019coherence}. The purpose of this work is to connect the average success probability of \texttt{GGSA} with the coherence fraction $ f_{c}(\rho) $. In the limiting case, the equal superposition state achieves the highest success probability. However, it is important to note that $f_{c}(\rho)$ is neither an entanglement measure nor a coherence measure of $\rho$. Only under specific constraints does it turn out to be a coherence measure. Our findings explain why the Hadamard gate is typically used to generate the equal superposition state in \texttt{GSA}, rather than an arbitrary quantum unitary gate.

\section{Generalized Grover Search Algorithm}
Grover's search algorithm (\texttt{GSA}) is a quantum algorithm designed for unstructured search tasks~\cite{grover1996fast, PhysRevLett.79.325, PhysRevLett.80.4329}. It efficiently locates $r$ elements within a dataset of size $N$ by searching through the indices. \texttt{GSA} achieves a quadratic speedup compared to classical search algorithms, reducing the computational complexity required for large datasets. For convenience, assume that the size of the dataset is $N=2^{n}$. Consider a function $f(x):\{0,1\}^{n}\rightarrow\{0,1\}$, where $f(x)=1$ if $x$ is a solution to the search problem, and $f(x)=0$ otherwise.

The implementation of \texttt{GSA} involves three main steps: (i) Apply the Hadamard gate $H^{\otimes n}$ to the input state $|0^{n}\rangle$ of the register to obtain $|\eta\rangle:=\sum_{x=0}^{N-1}|x\rangle/\sqrt{N}$; (ii) Perform the Grover operator $\mathcal{G}$ approximately $O(\sqrt{N})$ times; (iii) Measure the final state. The Grover operator $\mathcal{G}$ consists of two quantum subroutines: the quantum oracle $\mathcal{O}$ and the diffusion transform. The quantum oracle $\mathcal{O}$ rotates the phase of the marked states by $\pi$ radians. It corresponds to a unitary operation on the computational basis states of the register $|x\rangle$, and the ancilla $|q\rangle$: $\mathcal{O}|x\rangle|q\rangle=|x\rangle|q\oplus f(x)\rangle$. The ancilla qubit is initially set to the state $|-\rangle_q=(|0\rangle-|1\rangle)/\sqrt{2}$. With this choice, the oracle $\mathcal{O}$ can be described as $\mathcal{O}|x\rangle|-\rangle_q=(-1)^{f(x)}|x\rangle|-\rangle_q$. This implies that $|x\rangle$ is flipped if $f(x) =1$, and remains unchanged otherwise. The diffusion transform $\mathcal{D}=H^{\otimes n}(2|0^{n}\rangle\langle0^{n}|-I^{\otimes n})|H^{\otimes n}=2|\eta\rangle\langle\eta|-I^{\otimes n}$ applies to the state in the register.

\begin{figure}[h]
\centering
\includegraphics[width=3.4in]{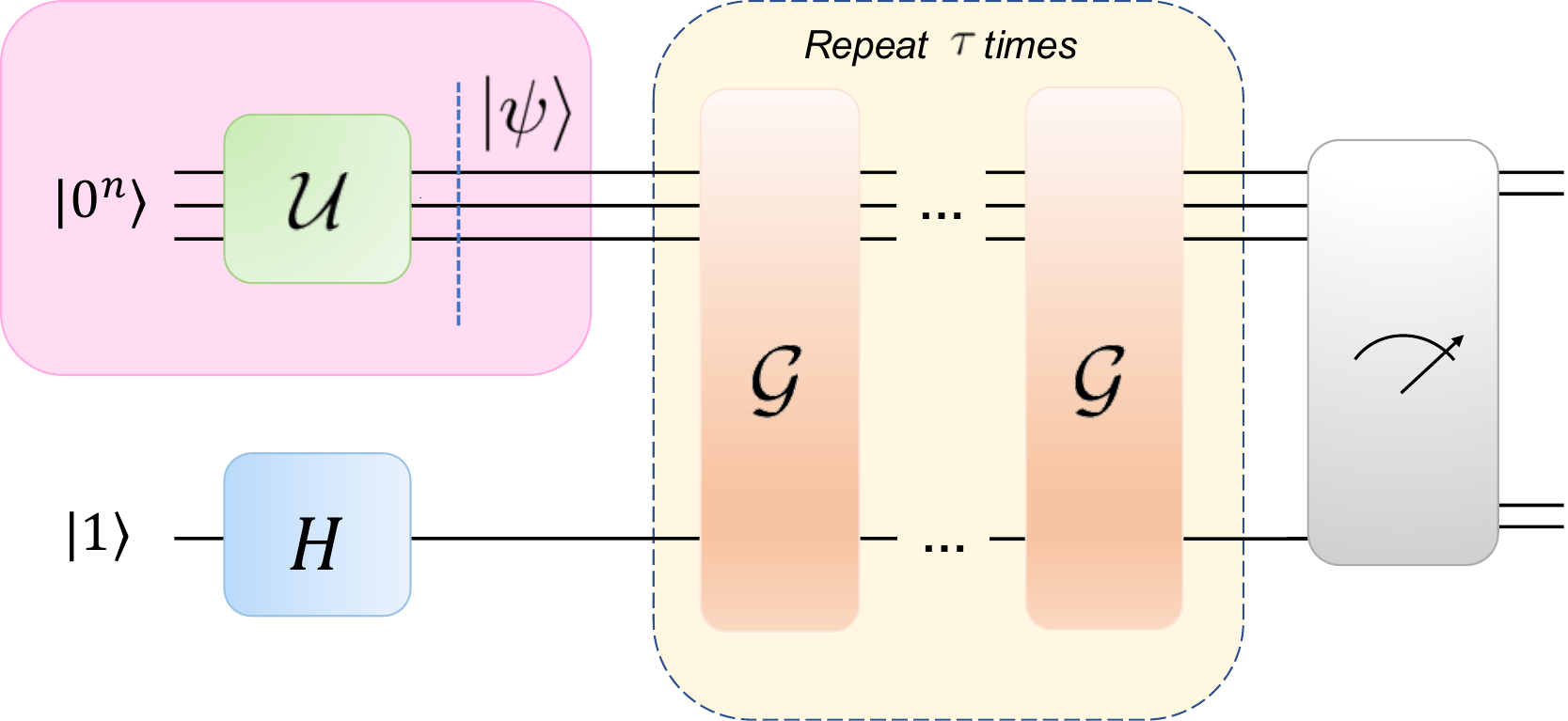}
\caption{\textbf{Quantum Circuit for \texttt{GGSA}.} An arbitrary unitary quantum gate $\mathcal{U}$ is applied to the input state $|0^{n}\rangle$ of the register, while the Hadamard gate $H$ is applied to the ancilla qubit $|1\rangle_q$. The initial state is $|\psi\rangle:=\mathcal{U}|0^{n}\rangle=\sum_{x=0}^{N-1}a_x|x\rangle$, where $a_x$ is the amplitude of $|x\rangle$. The resulting state is $|\psi\rangle|-\rangle$. Subsequently, the Grover operator $\mathcal{G}$ is applied $\tau$ times, and the final state is measured on the computational basis.}
\label{fig:GGSA}
\end{figure}

We consider a generalized Grover search algorithm (\texttt{GGSA}) detailed in Algo.~\ref{alg:GGSA}, with its corresponding circuit depicted in FIG.~\ref{fig:GGSA}. In contrast to the Hadamard gate used in \texttt{GSA}, \texttt{GGSA} implements an arbitrary unitary gate $\mathcal{U}$ before the Grover operator $\mathcal{G}$ to obtain the pure initial state $|\psi\rangle$. Additionally, it provides a direct generalization to the case in which the initial state is a mixed state $\rho$. We are interested in how the initial state impacts the algorithm's search performance and which intrinsic quantum properties contribute to the advantages of \texttt{GGSA}.

\begin{algorithm}[H]
\caption{\texttt{GGSA}}
Consider an oracle $\mathcal{O}$, a register of $n$ qubits initially in the state $|0^n\rangle$, and an ancilla qubit $|1\rangle_q$.
\begin{itemize}
    \item [1.] Apply an arbitrary unitary quantum gate $\mathcal{U}$ to the initial state $|0^{n}\rangle$ of the register, and the Hadamard gate $H$ to the ancilla qubit initially in state $|1\rangle_q$. The resulting state becomes $|\psi\rangle|-\rangle$,
    where $|\psi\rangle:=\sum_{x=0}^{N-1}a_x|x\rangle$.

    \item [2.] Perform the Grover iteration $\mathcal{G}=\mathcal{D}\mathcal{O}$ $\tau$ times, where $\mathcal{D}=H^{\otimes n}(2|0^{n}\rangle\langle0^{ n}-I^{\otimes n})|H^{\otimes n}$ represents the diffusion transform acting on the register state, and $\mathcal{O}|x\rangle|-\rangle_q=(-1)^{f(x)}|x\rangle|-\rangle_q$ denotes the oracle.

     \item [3.] Measure the resulting state.
\end{itemize}
\label{alg:GGSA}
\end{algorithm}

\section{The average success probability of \texttt{GGSA} and the coherence fraction}
\begin{figure}[h]
\centering
\includegraphics[width=3.4in]{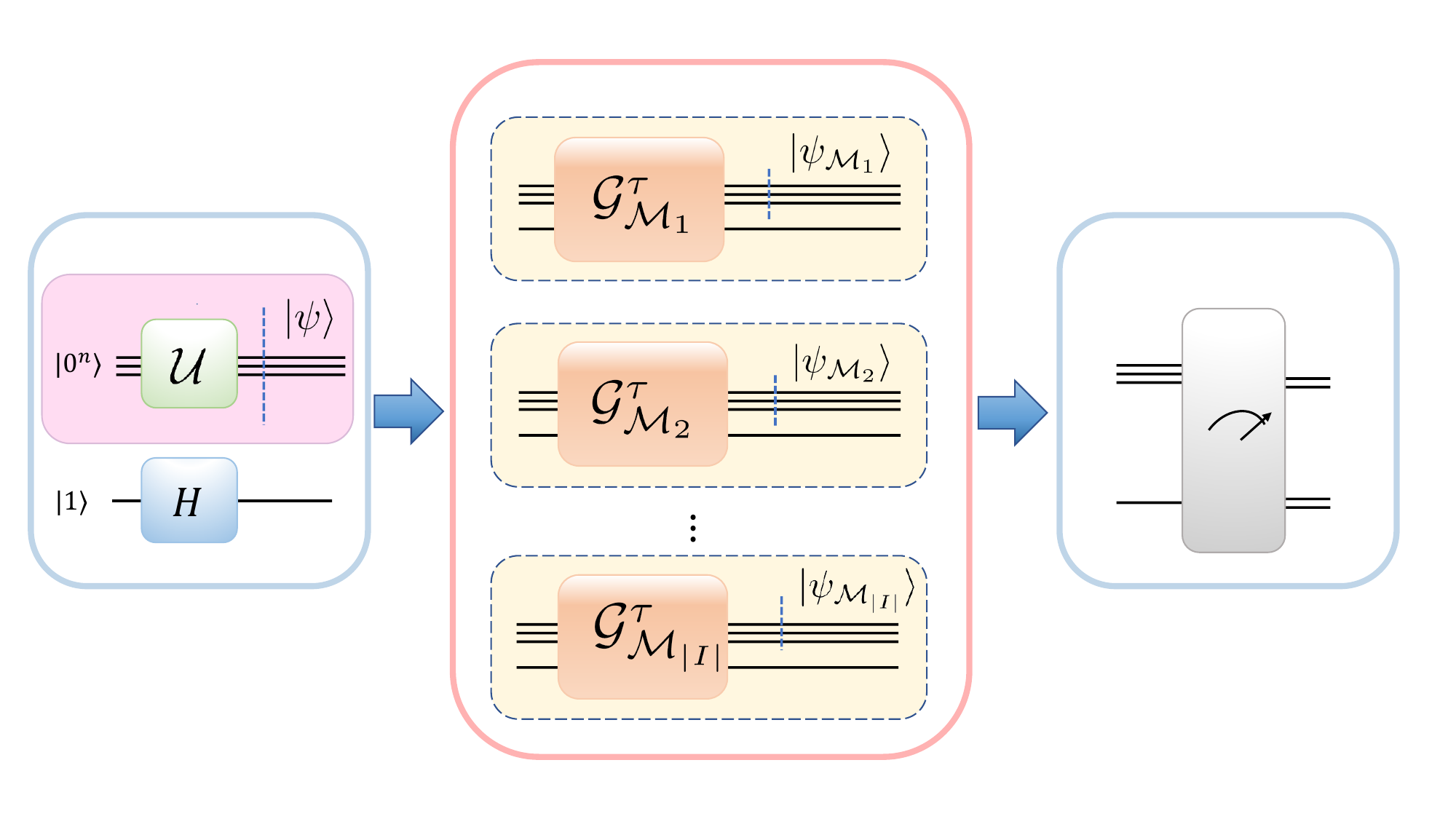}
\caption{
\textbf{Grover operator $\mathcal{G}_{\mathcal{M}_{i}}$.}
The algorithm is depicted in three outer boxes, each representing a different step. The first box initializes the state as $|\psi\rangle$. In the second box, the Grover operator $\mathcal{G}_{\mathcal{M}_{i}}$ is applied, resulting in the final state $|\psi_{\mathcal{M}_{i}}\rangle$. The small boxes inside signify the selections of a specific marked state each time the algorithm proceeds. The final box represents the measurement.}
\label{fig:GGSAreal}
\end{figure}

We investigate the performance of \texttt{GGSA} in terms of average success probability. Given $r$ marked states, the number of possible sets is represented by the binomial coefficient $\binom{N}{r}$~\cite{PhysRevA.68.022326, PhysRevA.71.042320}. Denote all possible sets of marked states as $\{\mathcal{M}_{i}\}_{i\in I}$, with $I$ being the indicator set encompassing all possible sets. The cardinality $|I|$ signifies the total number of sets, i.e., $|I|=\binom{N}{r}$. The oracle $\mathcal{O}|x\rangle|-\rangle_q=(-1)^{f(x)}|x\rangle|-\rangle_q$ in the Grover operator $\mathcal{G}$ is crucial for marking states. Consequently, in the second step, $\mathcal{G}$ is replaced by $\mathcal{G}_{\mathcal{M}_{i}}$ tailored for each set $\mathcal{M}_{i}$. After applying the Grover operator $\mathcal{G}_{\mathcal{M}_{i}}$ $\tau$ times to the state $|\psi\rangle$, we obtain $|\psi_{\mathcal{M}_{i}}\rangle$, as demonstrated in FIG.~\ref{fig:GGSAreal}. Each set $\{\mathcal{M}_{i}\}_{i\in I}$ corresponds to an operator $\mathcal{G}_{\mathcal{M}_{i}}$ and a specific final state $|\psi_{\mathcal{M}_{i}}\rangle$. Once a set is chosen, both the operator and the resulting final state are fixed.

The variability in the second step of Algo.~\ref{alg:GGSA} depends on the selection of marked states, influencing the diversity of possible outcomes. Let $P_{succ}^{\mathcal{M}_{i}}(|\psi\rangle, \tau)$ denote the success probability for the set $\mathcal{M}_{i}$. For any initial state $|\psi\rangle$, the average success probability $P_{\text{ave}}(|\psi\rangle, \tau)$ of \texttt{GGSA} over all possible sets after $\tau$ iterations is given by
\begin{equation}\label{III:eq1}
P_{\text{ave}}(|\psi\rangle, \tau):= \frac{1}{|I|}
\sum_{i}P_{succ}^{\mathcal{M}_{i}}(|\psi\rangle, \tau),
\end{equation}
where $|I|$ represents the total number of possible sets.

\begin{theorem}
For an initial state $|\psi\rangle$, the average success probability (\ref{III:eq1}) of \texttt{GGSA} is given by the following formula,
\begin{equation}\label{ave}
\begin{aligned}
P_{\text{ave}}(|\psi\rangle, \tau)
=\frac{N\sin^2\vartheta-r}{N-1}f_{c}(|\psi\rangle)
+\frac{r-\sin^2\vartheta}{N-1},
\end{aligned}
\end{equation}
where $\vartheta:= \theta(\tau+1/2)$, $\theta:=\arccos{(1-2r/N)}$ and $r$ is the number of marked elements in an unsorted database of size $N$.
\end{theorem}

The proof of Theorem 1 is presented in Appendix A. From Eq.~(\ref{ave}), we see that the average success probability $P_{\text{ave}}(|\psi\rangle, \tau)$ is a function of variables $\tau$ and the quantity $f_{c}(|\psi\rangle)$, where $f_{c}(|\psi\rangle)$ is the coherence fraction of the initial state defined by the fidelity between $|\psi\rangle$ and the equal superposition state $|\eta\rangle=\sum_{x=0}^{N-1}|x\rangle/\sqrt{N}$.

\begin{figure}[h]
\centering
\includegraphics[width=3in]{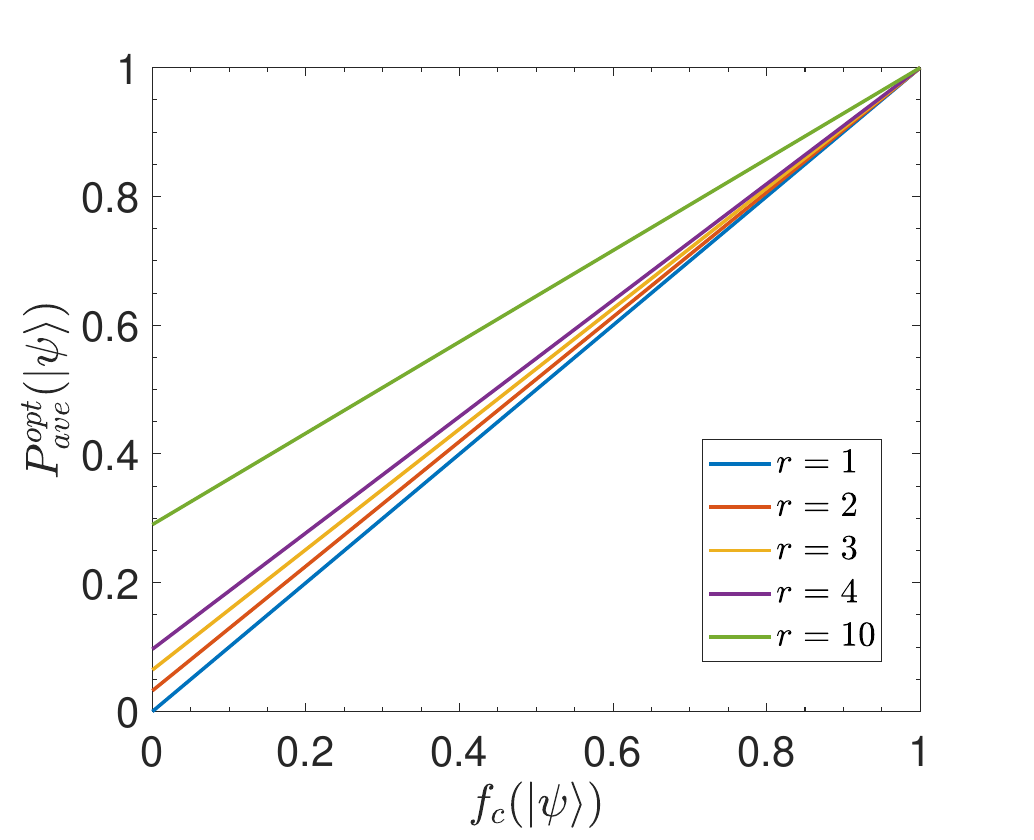}
\caption{\textbf{The optimal average success probability as a function of the coherence fraction.} We set $N=2^5$ and the plot follows the Eq.~(\ref{P_{opt}}).}
\label{fig:case0}
\end{figure}

For an optimal number of iterations $\tau_{opt}=\left\lfloor\frac{\pi}{4} \sqrt{\frac{N}{r}}\right\rfloor$ \cite{PhysRevLett.80.4329}, the optimal average success probability after performing $\tau_{opt}$ iterations is
\begin{align}\label{P_{opt}}
    P_{\text{ave}}^{\mathrm{opt}}(|\psi\rangle)=\frac{N-r}{N-1}f_{c}(|\psi\rangle)+\frac{r-1}{N-1},
\end{align}
where $\left\lfloor \cdot \right\rfloor$ is the floor function. Based on Eq.~(\ref{P_{opt}}), the optimal average success probability $P_{\text{ave}}^{\mathrm{opt}}(|\psi\rangle)$ is completely characterized by the coherence fraction $f_{c}(|\psi\rangle)$. FIG.~\ref{fig:case0} shows the relation between the optimal average success probability and the coherence fraction for $N=2^5$ and $r=1,2,3,4,10$. Note that $P_{\text{ave}}^{\mathrm{opt}}(|\psi\rangle)\in[(r-1)/(N-1),1]$ since $f_{c}(|\psi\rangle)\in[0,1]$. The lower bound of $P_{\text{ave}}^{\mathrm{opt}}(|\psi\rangle)$ is saturated when the initial state is $|\psi\rangle =\sum_{x=0}^{N-1}(-1)^{x}|x\rangle/\sqrt{N}=|-\rangle^{\otimes n}$. The upper bound of $P_{\text{ave}}^{\mathrm{opt}}(|\psi\rangle)$ is achieved when the initial state is $|\psi\rangle=\sum_{x=0}^{N-1}|x\rangle/\sqrt{N}=|+\rangle^{\otimes n}=|\eta\rangle$, which corresponds to the original Grover search algorithm \cite{PhysRevLett.80.4329}. It indicates that the Hadamard gate achieves the best performance in terms of maximum success probability. It is important to note that our previous analysis focused on the pure states. Similar results for mixed states are presented in Appendix B.

We have shown that the maximum success probability of \texttt{GGSA} is determined exclusively by the coherence fraction of the initial state. This discovery establishes a connection between the success probability and the coherence fraction. Here the coherence fraction is similar to the fully entangled fraction, which measures the fidelity of optimal quantum teleportation ~\cite{PhysRevA.54.3824, PhysRevA.54.1838, PhysRevA.60.1888, PhysRevA.62.012311, PhysRevA.66.012301, PhysRevA.91.012310}.

To explore the full potential of quantum computing, it is vital to understand which quantum properties offer the computational advantage. This insight will also facilitate the design of future quantum algorithms~\cite{jozsa2003role,ahnefeld2022coherence,PhysRevA.95.032307,PhysRevA.65.062312,PhysRevA.93.012111,anand2016coherenceentanglementmonogamydiscrete,liu2019coherence}. Various quantum resources, such as entanglement~\cite{RevModPhys.81.865, guhne2009entanglement}, coherence~\cite{gour2008resource,marvian2014extending,PhysRevLett.113.140401,PhysRevLett.118.020403,RevModPhys.89.041003,PhysRevLett.120.230504,hu2018quantum,PhysRevLett.125.060404, wu2021experimental, yang2023device} and discord~\cite{henderson2001classical, PhysRevLett.88.017901,PhysRevLett.100.050502}, have played significant roles in quantum information processing. Here, we emphasize that the coherence fraction $f_{c}(\rho)=F(|\eta\rangle, \rho)$ is neither an entanglement measure nor a coherence measure of $\rho$, and only under specific constraint does it turn out to be a coherence measure, see detailed proof in Appendix C.

\section{Examples}
We consider a special case in which the unitary quantum gate we applied in \texttt{GGSA} is a product of arbitrary local operations, $\mathcal{U}=\Big(\mathcal{U}(\alpha, \beta, \theta)\Big)^{\otimes{n}}$. The unitary gate applied to each qubit in the register is
\begin{align}
    \mathcal{U}(\alpha, \beta, \theta)=\begin{pmatrix}
 e^{i\alpha}\cos\theta & e^{-i\beta}\sin\theta \\ e^{i\beta}\sin\theta & -e^{-i\alpha}\cos\theta
\end{pmatrix},
\end{align}
where $\alpha$ and $\beta$ are parameters, $\theta \in [0, \pi/2]$. Especially, $\mathcal{U}(0, 0, \pi/4)$ is the Hadamard gate. Apply the unitary gate to the single qubit $|0^{n}\rangle$ and obtain the initial state
\begin{equation}
|\psi(\alpha, \beta, \theta)\rangle=\sum_{j=0}^{N-1}(e^{i\alpha}\cos\theta)^{k_{j}}(e^{i\beta}\sin\theta)^{n-k_{j}}|j\rangle.
\end{equation}
where $N=2^n$ and $k_{j}$ is the number of zero numbers in binary representation $j=j_{1}j_{2}\cdots j_{N}$. Thus, the coherence fraction of $|\psi(\alpha, \beta, \theta)\rangle$ is
\begin{equation}
f_{c}\Big(|\psi(\alpha, \beta, \theta)\rangle\Big)=\frac{1}{2^{n}}|(e^{i\alpha}\cos\theta+e^{i\beta}\sin\theta)^{n}|^{2}.
\end{equation}
For one marked state (i.e., $r=1$), the optimal average success probability is given by
\begin{equation}
P_{\text{ave}}^{\mathrm{opt}}\Big(|\psi(\alpha, \beta, \theta)\rangle\Big)
=f_{c}\Big(|\psi(\alpha, \beta, \theta)\rangle\Big).   \end{equation}
We consider two examples:
\begin{itemize}
\item[(a).]
$P_{\text{ave}}^{\mathrm{opt}}\Big(|\psi(\alpha, \beta, \pi/4)\rangle\Big)=\frac{1}{4^{n}}|(e^{i\alpha}+e^{i\beta})^{n}|^{2}$; \item[(b).]$P_{\text{ave}}^{\mathrm{opt}}\Big(|\psi(0, 0, \theta)\rangle\Big)=\frac{1}{2^{n}}(\cos\theta+\sin\theta)^{2n}$.
\end{itemize}
The link between parameters ($\alpha$, $\beta$, $\theta$) and the optimal average success probability $P_{\text{ave}}^{\mathrm{opt}}\Big(|\psi(\alpha, \beta, \theta)\rangle\Big)$ of these two examples are shown in FIG.~\ref{fig:case}. See more details in Appendix D.

\begin{figure}[h]
\centering
\includegraphics[width=3.5in]{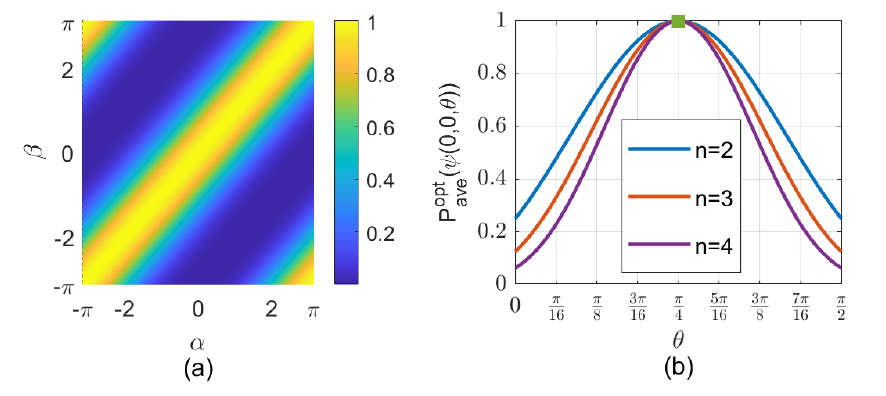}
\caption{\textbf{The link between parameters ($\alpha$, $\beta$, $\theta$) and the optimal average success probability $P_{\text{ave}}^{\mathrm{opt}}\Big(|\psi(\alpha, \beta, \theta)\rangle\Big)$.} \textbf{Example (a).} Fixed the parameter $\theta=\pi/4$, the success probability $P_{\text{ave}}^{\mathrm{opt}}\Big(|\psi(\alpha, \beta, \pi/4)\rangle\Big)=1$ when $\alpha=\beta+2k\pi$, $(k=0, \pm 1, \cdots)$. Here is the case of $n=2$. \textbf{Example (b).} Fixed parameters $\alpha=\beta=0$, the upper bound of $P_{\text{ave}}^{\mathrm{opt}}\Big(|\psi(0, 0, \theta)\rangle\Big)=1$ is achieved when $\theta=\pi/4$. Three lines in the figure demonstrate 2, 3, and 4 qubits from top to bottom.}
\label{fig:case}
\end{figure}

\section{Quantum Minimization Algorithm}
We further discuss the role of coherence fraction in the D\"urr and H\o{}yer's quantum minimization algorithm (\texttt{QMA})~\cite{durr1996quantum}. The \texttt{QMA} offers a groundbreaking way to tackle complex issues in artificial intelligence. It is an important subroutine in unsupervised and supervised quantum machine learning~\cite{biamonte2017quantum,wittek2014quantum, PhysRevLett.113.130503, schuld2015introduction}. In particular, \texttt{QMA} can be integrated into quantum versions of $k$-means and $k$-median algorithms for unsupervised learning, as well as quantum support vector machines for supervised learning.
The \texttt{QMA} solves general optimization problems by performing the \texttt{GSA} multiple times.

\begin{algorithm}[H]
\caption{\texttt{GQMA}}
1. Randomly choose an input $x_{1}$ and set $d_{1}= f(x_{1})$.

2. Repeat the following steps until convergence is attained:

\begin{itemize}
  \item [(1)] Initialize the first register with $n$ input qubits and the second register with $m$ qubits to hold the threshold value $d_i$.
  \item [(2)] Apply an arbitrary unitary quantum gate $\mathcal{U}$ to the first register to prepare an arbitrary state. Initialize the memory as $|\psi\rangle\otimes |d_{i}\rangle$. Mark each item $x$ where $f(x)<d_{1}$.
  \item [(3)] Apply the Grover operator $\mathcal{G}$ approximately $\tau_i$ times. Let the resulting state be  $|x\rangle$, and evaluate the function value $d = f(x)$.
  \item [(4)] Update the threshold: If $d < d_i$, set $x_{i+1} = x$  and $d_{i+1} = d$. Otherwise, retain the previous values by setting $x_{i+1} = x_i$ and $d_{i+1} = d_i$.
\end{itemize}
3. Return $x$.
\label{alg:GQMA}
\end{algorithm}

\begin{figure}[h]
\centering
\includegraphics[width=3.4in]{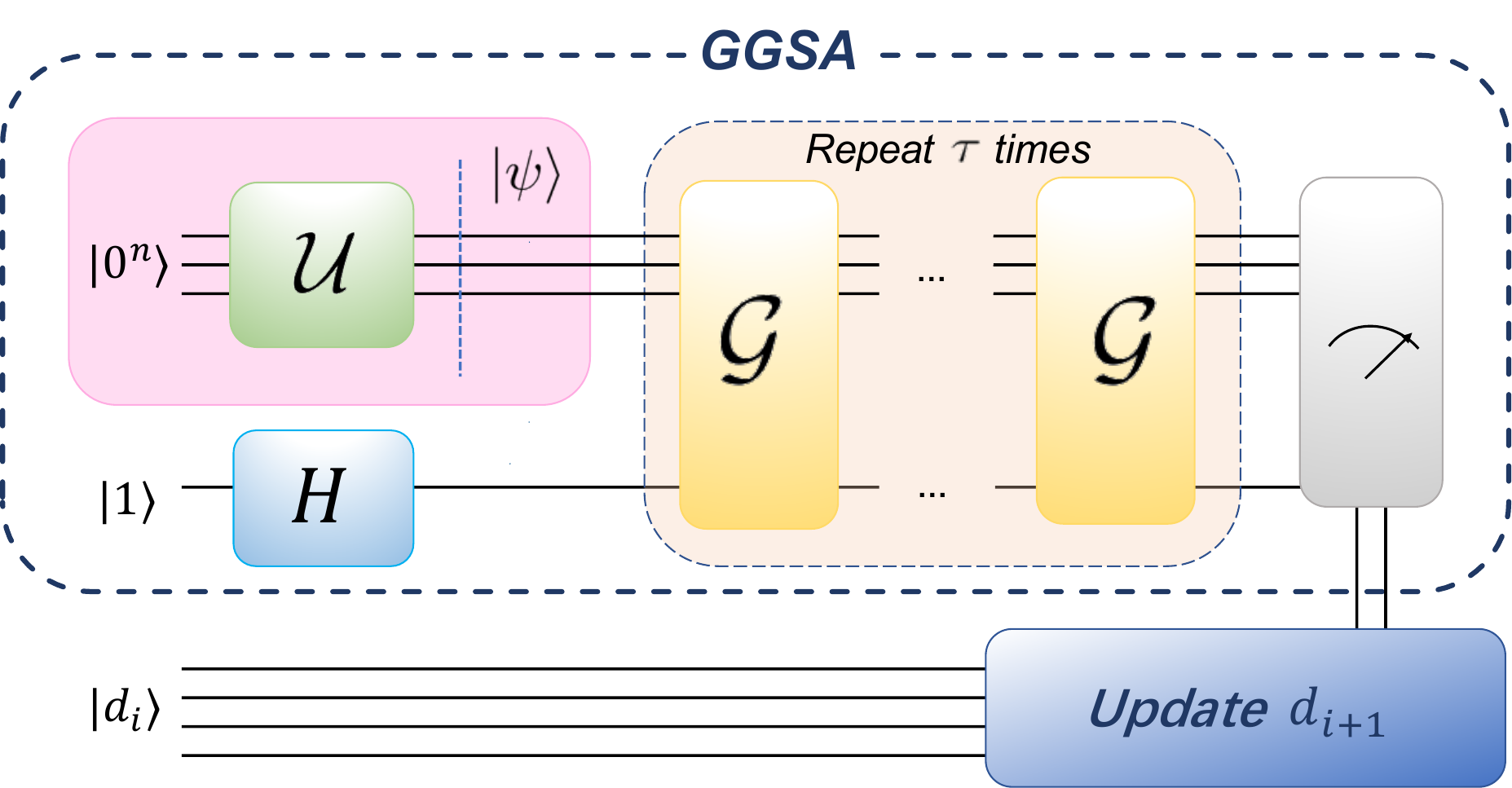}
\caption{\textbf{Quantum Circuit for \texttt{GQMA}.}
The main subroutine of \texttt{GQMA} is the \texttt{GGSA}, as indicated by the outermost dashed box. In contrast to \texttt{GGSA}, here it involves three registers.
Initialize the first register with \( n \) input qubits and the second register with \( m \) qubits to store the threshold \( d_i \). Apply an arbitrary unitary quantum gate \( \mathcal{U} \) to the first register to prepare an arbitrary state. Set the initial state to \( |\psi\rangle \otimes |d_i\rangle \). Mark all items \( x \) for which \( f(x) < d_1 \). Execute the Grover operator \( \mathcal{G} \) for approximately \( \tau_i \) iterations. Measure the output state \( |x\rangle \) and compute the function value \( d = f(x) \). Update the threshold: if \( d < d_i \), then set \( x_{i+1} = x \) and \( d_{i+1} = d \). Otherwise, keep the previous values by setting \( x_{i+1} = x_i \) and \( d_{i+1} = d_i \).
}
\label{fig:GQMA}
\end{figure}

We here present a generalized version of the quantum minimization algorithm (\texttt{GQMA}). The main subroutine is a generalization of \texttt{GSA} called quantum exponential searching algorithm~\cite{boyer1998tight}. Consider an unsorted table of $N$ items, ${0, 1, \cdots, N-1}$, each holding a value from an ordered set. The algorithm requires the objective function
$f(x):\{0,1\}^{n} \rightarrow \mathbb{R}$. The goal is to find the index $x$ such that $f(x)$ is minimized, theoretically identifying the global optimum $x_{\min}$ at the end. Building on \texttt{QMA}, we consider a search using \texttt{GQMA} for one minimum among $r$ marked states within a space of $N=2^n$ computational basis states, where $n$ is the number of qubits in the first register. Instead of $|+\rangle^{\otimes n}$, we apply an arbitrary unitary quantum gate $\mathcal{U}$ to $|0\rangle^{\otimes n}$ to obtain an arbitrary pure state. The detailed steps of \texttt{GQMA} are explained below in Algo.~\ref{alg:GQMA}, and the corresponding quantum circuit is presented in FIG.~\ref{fig:GQMA}.

To analyze the probability of success for the aforementioned scenario, assume that the \texttt{GQMA} algorithm runs sufficiently long to find the global minimum. The algorithm finding the global minimum $ x_{\min} $ implies that $r=1$. Suppose that convergence is attained after $ s $ iterations. In the limit of a large search space and with an arbitrary initial state $ |\psi\rangle $, the global minimum $ x_{\min} $ can be determined after $ \tau_{s} $ iterations, with the success probability given by
\begin{equation}
\begin{aligned}
P_{\text{ave}}(|\psi\rangle, \tau_{s})
=\frac{N\sin^2\vartheta-1}{N-1}f_{c}(|\psi\rangle)
+\frac{1-\sin^2\vartheta}{N-1},
\end{aligned}
\end{equation}
where $\vartheta:= \theta(\tau_{s}+1/2)$ and $\theta:=\arccos{(1-2/N)}$. After $\tau_{s}^{opt}=\left\lfloor\pi\sqrt{N}/4 \right\rfloor$ iterations, the success probability of obtaining the correct measurement is upper-bounded by
\begin{equation}
   P_{\text{ave}}^{\mathrm{opt}}(|\psi\rangle)=f_{c}(|\psi\rangle).
\end{equation}
Thus, there is a monotonic relationship between the coherence fraction $f_{c}(|\psi\rangle)$ and the optimal average success probability.

\section{Discussions}
We have systematically explored the relationship between the intrinsic quantum properties of the initial quantum states and the efficacy of quantum algorithms. Our analysis reveals a clear connection between the success probability and the coherence fraction of the initial state. Notably, the maximum coherence fraction corresponds to the highest success probability in the Grover search algorithm. These findings emphasize the crucial significance of the equal superposition state in quantum computation, which implies why the Hadamard gate is frequently employed to generate such states in the design of quantum algorithms.

While there have been exciting discoveries in quantum algorithms, it is important to note that not all quantum algorithms have been found to outperform their classical counterparts~\cite{tang2019quantum,tang2021quantum,shao2024simulating}. However, Shor's algorithm and Grover's search algorithm are two quantum algorithms that offer significant quantum advantages. In particular, Shor's factoring algorithm provides a super-polynomial speedup compared to all currently known classical factoring algorithms~\cite{365700}. On the other hand, Grover's search algorithm has been proven to exhibit a quadratic speedup compared to any classical algorithms~\cite{grover1996fast, PhysRevLett.79.325, bennett1997strengths, PhysRevA.60.2746}. Recent research has highlighted the crucial role of quantum coherence in achieving the advantage of Shor's factoring algorithm with a fixed structure~\cite{ahnefeld2022coherence}. These findings indicate that the quantum advantage of the Grover search algorithm is attributed to the coherence fraction of quantum states, rather than quantum coherence and entanglement. While our focus is primarily on the Grover search algorithm, these insights can be applied to other algorithms based on Grover's principles as well~\cite{Gilliam2021groveradaptive, ohno2024grover}. Our results provide a new perspective on understanding the quantum advantages of different quantum algorithms and may enable the design of algorithms that truly achieve quantum advantages.

\section*{Acknowledgements}
This work was supported by Fundamental Research Funds for the Central Universities, the National Natural Science Foundation of China (12371132, 12075159, 12171044, 12405006), and the specific research fund of the Innovation Platform for Academicians of Hainan Province.
Y. X. is supported by the National Research Foundation, Singapore, A*STAR Quantum Engineering Programme (NRF2021-QEP2-02-P03), A*STAR Central Research Funds, and Q.InC Strategic Research and Translational Thrust.

\appendix
\section{Proof of Theorem 1}

Consider a search space containing $N=2^n$ elements in which a subset of $r$ elements are marked. The initial state is given by in the basis $|x\rangle$,
\begin{equation}
\begin{aligned}
|\psi\rangle&=\sum_{x=0}^{N-1}a_x|x\rangle\\
&=\sum_{m\in\mathcal{M}}a_m|m\rangle+\sum_{u\in\mathcal{U}}a_u|u\rangle\\
&=\sqrt{P_0}|\phi_M\rangle+\sqrt{1-P_0}|\phi_U\rangle,
\end{aligned}
\end{equation}
where $|\phi_M\rangle=\frac{1}{\sqrt{P_0}}\sum_{m\in\mathcal{M}}a_m|m\rangle$, $|\phi_U\rangle=\frac{1}{\sqrt{1-P_0}}\sum_{u\in\mathcal{U}}a_u|u\rangle$, $P_0=\sum_{m\in\mathcal{M}}|a_m|^2$, $\mathcal{M}$ and $\mathcal{U}$ denote the marked and unmarked sets, respectively.
The equally superposed state can be correspondingly written as,
\begin{equation}
\begin{aligned}
|\eta\rangle&=\frac{1}{\sqrt{N}}\sum_{x=0}^{N-1}|x\rangle\\
&=\sqrt{\frac{r}{N}}|\eta_M\rangle+\sqrt{\frac{N-r}{N}}|\eta_U\rangle.
\end{aligned}
\end{equation}
\par
Using the Gram-Schmidt procedure we construct two orthogonal states,
\begin{equation}
\begin{aligned}
&|\psi_M\rangle=\frac{\sqrt{P_0}|\phi_M\rangle-\sqrt{r}\bar{a}_M|\eta_M\rangle}{\sqrt{P_0-r|\bar{a}_M|^2}},\\
&|\psi_U\rangle=\frac{\sqrt{1-P_0}|\phi_U\rangle-\sqrt{N-r}\bar{a}_U|\eta_U\rangle}{\sqrt{1-P_0-(N-r)|\bar{a}_M|^2}},
\end{aligned}
\end{equation}
where $\bar{a}_M=\frac{1}{r}\sum_{m\in\mathcal{M}}a_m$ and $\bar{a}_U=\frac{1}{N-r}\sum_{u\in\mathcal{U}}a_u$. The initial state can then be expressed as
\begin{equation}
\begin{aligned}
|\psi\rangle=&\sqrt{P_0-r|\bar{a}_M|^2}|\psi_M\rangle+\sqrt{N-r}\bar{a}_U|\eta_U\rangle\\
&+\sqrt{1-P_0-(N-r)|\bar{a}_U|^2}|\psi_U\rangle+\sqrt{r}\bar{a}_M|\eta_M\rangle,
\end{aligned}
\end{equation}
which is spanned spanned by $|\psi_M\rangle$, $|\eta_M\rangle$ and $|\psi_U\rangle$, $|\eta_U\rangle$. For convenience, we use the following vector representations,
\begin{align}
&\left|\psi_{M}\right\rangle \equiv\left(\begin{array}{c}1 \\ 0 \\ 0 \\ 0\end{array}\right),
~\left|\psi_{U}\right\rangle \equiv\left(\begin{array}{l}0 \\ 1 \\ 0 \\ 0\end{array}\right)\\
&\left|\eta_{U}\right\rangle \equiv\left(\begin{array}{l}0 \\ 0 \\ 1 \\ 0\end{array}\right),
~\left|\eta_{M}\right\rangle \equiv\left(\begin{array}{l}0 \\ 0 \\ 0 \\ 1\end{array}\right).
\end{align}
Thus the Grover iteration $\mathcal{G}$ acts as a linear transformation within a four-dimensional vector space. Its matrix representation $\mathcal{Q}$ is given by
\begin{equation}
\mathcal{Q}=
\begin{pmatrix}
1&  0& 0& 0\\
0& -1& 0& 0\\
0&  0& \cos\theta& -\sin\theta\\
0&  0& \sin\theta& \cos\theta
\end{pmatrix}.
\end{equation}

After $\tau$ iterations the state is of the form,
\begin{equation}
\begin{aligned}
|\psi(\tau)\rangle&=\mathcal{Q}^\tau|\psi\rangle\\
&=\sqrt{P_0-r|\bar{a}_M|^2}|\psi_M\rangle\\
&+(-1)^\tau\sqrt{1-P_0-(N-r)|\bar{a}_U|^2}|\psi_U\rangle\\
&+(\sqrt{N-r}\bar{a}_U\cos(\theta\tau)-\sqrt{r}\bar{a}_M\sin(\theta\tau))|\eta_U\rangle\\
&+(\sqrt{N-r}\bar{a}_U\sin(\theta\tau)+\sqrt{r}\bar{a}_M\cos(\theta\tau))|\eta_M\rangle,
\end{aligned}
\end{equation}
where $\cos\theta=1-\frac{2r}{N}$.
The success probability $P_{succ}(\tau)$ of a measurement is given via the projection of $|\psi(\tau)\rangle$ on the plane of marked states:
\begin{equation}
\begin{aligned}
P_{succ}(\tau)&=P_0+\frac{1}{2}[(N-r)|\bar{a}_U|^2-r|\bar{a}_M|^2]\\
&-\frac{1}{2}[(N-r)|\bar{a}_U|^2-r|\bar{a}_M|^2]\cos(2\theta\tau)\\
&+\frac{1}{2}\sqrt{r(N-r)}(\bar{a}_U^{*}\bar{a}_M+\bar{a}_U\bar{a}_M^{*})\sin(2\theta\tau).
\end{aligned}
\end{equation}
This provides an upper bound $P_{max}$ of the success probability $P_{succ}(\tau)$,
\begin{equation}
P_{max}=P_{0}+(N-r)|\bar{a}_U|^2.
\end{equation}

The average of $P_0$ over all possible choices of the $r$ marked states is
given by
\begin{equation}
\begin{aligned}
\langle P_0 \rangle=\langle \sum_{m}|a_m|^2 \rangle=\frac{r}{N}\sum_{j=0}^{N-1}|a_j|^2=\frac{r}{N}.
\end{aligned}
\end{equation}
Since
\begin{equation}
\begin{aligned}
&\langle |\bar{a}_M|^2 \rangle=\frac{r-1}{r(N-r)}f_{c}(|\psi\rangle)+\frac{N-r}{Nr(N-1)},\\
&\langle |\bar{a}_U|^2 \rangle=\frac{N-r-1}{(N-1)(N-r)}f_{c}(|\psi\rangle)+\frac{r}{N(N-1)(N-r)},\\
&\langle \bar{a}_U^{*}\bar{a}_M \rangle = \langle \bar{a}_U\bar{a}_M^{*} \rangle=\frac{1}{N-1}f_{c}(|\psi\rangle)+\frac{1}{N(N-1)},
\end{aligned}
\end{equation}
where $f_{c}(|\psi\rangle)=F(|\eta\rangle, |\psi\rangle)$ is the Uhlmann's fidelity $F$ between quantum states $|\psi\rangle $ and $ |\eta\rangle $,
we obtain the average of $P_{succ}(\tau)$ over all possible choices of the set of $r$ marked states,
\begin{align}
P_{\text{ave}}(|\psi\rangle, \tau)&=\frac{Nf_{c}(|\psi\rangle)-1}{N-1}\sin^2[\theta(\tau+1/2)]\nonumber\\
&-\frac{rf_{c}(|\psi\rangle)}{N-1}+\frac{r}{N-1}.
\end{align}

It is clear that $P_{\text{ave}}(|\psi\rangle, \tau)$ depends on the coherence fraction $f_{c}(|\psi\rangle)$ and the quantity $\sin^2[\theta(\tau+1/2)]$. For a fixed $f_{c}(|\psi\rangle)$, the average success probability $P_{\text{ave}}(|\psi\rangle, \tau)$ is upper bounded by the following inequality,
\begin{equation}
P_{\text{ave}}(|\psi\rangle, \tau)\leq P_{\text{ave}}^{\mathrm{opt}}(|\psi\rangle)=\frac{N-r}{N-1}f_{c}(|\psi\rangle)+\frac{r-1}{N-1},
\end{equation}
where we use the fact that $\sin^2[\theta(\tau+1/2)]\in[0,1]$. The equality is attained at $\theta(\tau+1/2)\approx \frac{\pi}{2}$. Thus, we obtain $\tau=\left\lfloor\frac{\pi}{2\theta}-\frac{1}{2}\right\rfloor$. As $\cos\theta=1-\frac{2r}{N}$, namely, $\sin\theta=\frac{2}{N}\sqrt{r(N-r)}$, for large $N$, $\theta \approx \frac{2}{N}\sqrt{r(N-r)}$. As a result, we get $\tau=\left\lfloor\frac{N\pi\sqrt{r(N-r)}}{4r(N-r)}\right\rfloor$. For $r \ll N$, $\tau=\left\lfloor\frac{N\pi\sqrt{rN}}{4rN}\right\rfloor=\left\lfloor\frac{\pi}{4} \sqrt{\frac{N}{r}}\right\rfloor$.
The optimal number of iterations is $\tau_{opt}=\left\lfloor\frac{\pi}{4} \sqrt{\frac{N}{r}}\right\rfloor$ \cite{PhysRevLett.80.4329}.

\section{The average success probability of \texttt{GGSA} for mixed initial states}

Assume the initial state of the register is an arbitrary $n$-qubit mixed state,
\begin{equation}
\rho=\sum_{\mu}p_{\mu}|\psi_{\mu}\rangle\langle\psi_{\mu}|,
\end{equation}
where $\sum_{\mu}p_{\mu}=1$, $|\psi_{\mu}\rangle$ has the form,
\begin{equation}
|\psi_{\mu}\rangle=\sum_{i=0}^{N-1}a_{\mu i}|i\rangle.
\end{equation}
After $\tau$ Grover iterations, the pure state $|\psi_{\mu}\rangle$ is changed to $\mathcal{G}^{\tau}|\psi_{\mu}\rangle$, with the  corresponding density operator given by
\begin{equation}
\rho(\tau)=\sum_{\mu}p_{\mu}\mathcal{G}^{\tau}|\psi_{\mu}\rangle\langle\psi_{\mu}|\mathcal{G}^{\tau\dagger}.
\end{equation}
Averaging the probability of success over all possible choices of the set of $r$ ($r \ll N$) marked states, we have
\begin{equation}
\begin{aligned}
P_{\text{ave}}(\rho, \tau)
&=\frac{N\sin^2\vartheta-r}{N-1}\sum_{\mu}p_{\mu}f_{c}(|\psi_{\mu}\rangle)
+\frac{r-\sin^2\vartheta}{N-1}\\
&=\frac{N\sin^2\vartheta-r}{N-1}f_{c}(\rho)
+\frac{r-\sin^2\vartheta}{N-1}.
\end{aligned}
\end{equation}
where $\vartheta:= \theta(\tau+1/2)$ and $\theta=\arccos{(1-2r/N)}$, $f_{c}(\rho)=\langle \eta | \rho | \eta \rangle$ is the coherence fraction of the initial state $\rho$.
Thus, after $\tau_{opt}=\left\lfloor\pi\sqrt{N}/4 \right\rfloor$ iterations, the upper bound of the average success probability is given as follows,
\begin{equation}
\begin{aligned}
P_{\text{ave}}^{\mathrm{opt}}(\rho)
&=\frac{N-r}{N-1}\sum_{\mu}p_{\mu}f_{c}(|\psi_{\mu}\rangle)+\frac{r-1}{N-1}\\
&=\frac{N-r}{N-1}f_{c}(\rho)+\frac{r-1}{N-1}.
\end{aligned}
\end{equation}
\section{About the measure properties of coherence fraction}
\subsection{Coherence fraction is not a measure in general}
We show that the coherence fraction $f_{c}(\rho)= F(|\eta\rangle, \rho)$ is neither an entanglement measure nor a coherence measure of $\rho$.

\textbf{(1)} Consider a separable state $|\phi_{0}\rangle=\frac{1}{\sqrt{2}}(|00\rangle+|10\rangle)=\frac{1}{\sqrt{2}}(|0\rangle+|1\rangle)\otimes |0\rangle$ and an entangled state $|\phi_{1}\rangle=\frac{1}{\sqrt{2}}(|00\rangle+|11\rangle)$. Obviously, $f_{c}(\phi_{0})=f_{c}(\phi_{1})=\frac{1}{2}$, i.e., $f_{c}$ cannot distinguish a separable state from an entangled one.

\textbf{(2)} Consider the incoherent state $\rho_{0}$ and the coherent state $\rho_{1}$,
\begin{align}
    \rho_{0}=
\begin{pmatrix}
\frac{1}{2}&   0\\\\
0&  \frac{1}{2}
\end{pmatrix},~
\rho_{1}=\begin{pmatrix}
\frac{1}{3}&   -\frac{1}{3}i\\\\
\frac{1}{3}i&  \frac{2}{3}
\end{pmatrix}.
\end{align}
We have $f_{c}(\rho_{0})=f_{c}(\rho_{1})=\frac{1}{2}$, namely,
$f_{c}$ cannot distinguish an incoherent state from a coherent one.

\subsection{Coherence fraction as a coherence measure under specific conditions}

For states $\rho$ such that $\langle i|\rho|j\rangle \geqslant 0$ for all $i$ and $j$, $C(\rho):=f_{c}(\rho)-1/N =F(|\eta\rangle, \rho)-1/N$ is just the $l_{1}$-norm of coherence~\cite{PhysRevLett.113.140401}. Therefore, in this case, $f_{c}(\rho)$ quantifies the coherence, up to a constant factor $1/N$. Moreover, the optimal average success probability can be rewritten as $P_{\text{ave}}^{\mathrm{opt}}(\rho)=(N-r)C(\rho)/(N-1)+r/N$.
\vspace{9pt}
\section{Detailed derivations for the Example}
Consider the following unitary quantum gate $\mathcal{U}$,
\begin{equation}
   \mathcal{U}=\Big(\mathcal{U}(\alpha, \beta, \theta)\Big)^{\otimes{n}},
\end{equation}
where the unitary gate $\mathcal{U}(\alpha, \beta, \theta)$ applied to each qubit in the register is given by
\begin{equation}
\mathcal{U}(\alpha, \beta, \theta)
=\begin{bmatrix}
    e^{i\alpha}\cos\theta & e^{-i\beta}\sin\theta \\
    e^{i\beta}\sin\theta & -e^{-i\alpha}\cos\theta
\end{bmatrix}
\end{equation}
with parameters $\alpha$, $\beta$ and $\theta \in [0, \pi/2]$. Especially, $\mathcal{U}(0, 0, \pi/4)$ is the Hadamard operator.

Applying the unitary gate to the single qubit state $|0\rangle$, we obtain
\begin{align}
    |\phi(\alpha, \beta, \theta)\rangle&=\mathcal{U}(\alpha, \beta, \theta)|0\rangle\\
    &=e^{i\alpha}\cos\theta|0\rangle+e^{i\beta}\sin\theta|1\rangle.
\end{align}
The initial state is then given by
\begin{align}
    |\psi(\alpha, \beta, \theta)\rangle&=|\phi(\alpha, \beta, \theta)\rangle^{\otimes n}\\
    &=\sum_{j=0}^{N-1}(e^{i\alpha}\cos\theta)^{k_{j}}(e^{i\beta}\sin\theta)^{n-k_{j}}|j\rangle,
\end{align}
where $N=2^n$ and $k_{j}$ is the number of zeros in the binary representation of $j=j_{1}j_{2}\cdots j_{N}$.
The coherence fraction of $|\psi(\alpha, \beta, \theta)\rangle$ is
\begin{equation}
f_{c}\Big(|\psi(\alpha, \beta, \theta)\rangle\Big)=\frac{1}{2^{n}}|(e^{i\alpha}\cos\theta+e^{i\beta}\sin\theta)^{n}|^{2}.
\end{equation}
The optimal average success probability after $\tau_{opt}$ iterations is given by
\begin{align}
    P_{\text{ave}}^{\mathrm{opt}}(|\psi\rangle)=\frac{N-r}{N-1}f_{c}(|\psi\rangle)+\frac{r-1}{N-1},
\end{align}
For one marked state (i.e., $r=1$), the optimal average success probability is given by
\begin{align}
P_{\text{ave}}^{\mathrm{opt}}\Big(|\psi(\alpha, \beta, \theta)\rangle\Big)
&=f_{c}\Big(|\psi(\alpha, \beta, \theta)\rangle\Big)\\
&=\frac{1}{2^{n}}|(e^{i\alpha}\cos\theta+e^{i\beta}\sin\theta)^{n}|^{2}.
\end{align}
In particular, we have
\begin{align}
    P_{\text{ave}}^{\mathrm{opt}}\Big(|\psi(\alpha, \beta, \pi/4)\rangle\Big)&=\frac{1}{4^{n}}|(e^{i\alpha}+e^{i\beta})^{n}|^{2},\\
    P_{\text{ave}}^{\mathrm{opt}}\Big(|\psi(0, 0, \theta)\rangle\Big)&=\frac{1}{2^{n}}(\cos\theta+\sin\theta)^{2n}.
\end{align}

\bibliographystyle{apsrev4-1}
\bibliography{CFG}

@article{bennett2000quantum,
  title={Quantum information and computation},
  author={Bennett, Charles H and DiVincenzo, David P},
  journal={Nature},
  volume={404},
  number={6775},
  pages={247--255},
  year={2000},
  publisher={Nature Publishing Group},
  doi = {10.1038/35005001},
  url = {https://doi.org/10.1038/35005001}
}

@INPROCEEDINGS{365700,
  author={Shor, P.W.},
  booktitle={Proceedings 35th Annual Symposium on Foundations of Computer Science}, 
  title={Algorithms for quantum computation: discrete logarithms and factoring}, 
  year={1994},
  volume={},
  number={},
  pages={124-134},
  doi={10.1109/SFCS.1994.365700}}

@inproceedings{grover1996fast,
  title={A fast quantum mechanical algorithm for database search},
  author={Grover, Lov K},
  booktitle={Proceedings of the twenty-eighth annual ACM symposium on Theory of computing},
  pages={212--219},
  year={1996},
  doi={10.1145/237814.237866}
}

@article{PhysRevLett.79.325,
  title = {Quantum Mechanics Helps in Searching for a Needle in a Haystack},
  author = {Grover, Lov K.},
  journal = {Phys. Rev. Lett.},
  volume = {79},
  issue = {2},
  pages = {325--328},
  numpages = {0},
  year = {1997},
  month = {Jul},
  publisher = {American Physical Society},
  doi = {10.1103/PhysRevLett.79.325},
  url = {https://link.aps.org/doi/10.1103/PhysRevLett.79.325}
}

@article{PhysRevLett.103.150502,
  title = {Quantum Algorithm for Linear Systems of Equations},
  author = {Harrow, Aram W. and Hassidim, Avinatan and Lloyd, Seth},
  journal = {Phys. Rev. Lett.},
  volume = {103},
  issue = {15},
  pages = {150502},
  numpages = {4},
  year = {2009},
  month = {Oct},
  publisher = {American Physical Society},
  doi = {10.1103/PhysRevLett.103.150502},
  url = {https://link.aps.org/doi/10.1103/PhysRevLett.103.150502}
}

@incollection{feynman2018simulating,
  title={Simulating physics with computers},
  author={Feynman, Richard P},
  booktitle={Feynman and computation},
  pages={133--153},
  year={2018},
  publisher={CRC Press},
  url = {https://doi.org/10.1007/BF02650179}
}

@article{RevModPhys.86.153,
  title = {Quantum simulation},
  author = {Georgescu, I. M. and Ashhab, S. and Nori, Franco},
  journal = {Rev. Mod. Phys.},
  volume = {86},
  issue = {1},
  pages = {153--185},
  numpages = {33},
  year = {2014},
  month = {Mar},
  publisher = {American Physical Society},
  doi = {10.1103/RevModPhys.86.153},
  url = {https://link.aps.org/doi/10.1103/RevModPhys.86.153}
}

@article{biamonte2017quantum,
  title={Quantum machine learning},
  author={Biamonte, Jacob and Wittek, Peter and Pancotti, Nicola and Rebentrost, Patrick and Wiebe, Nathan and Lloyd, Seth},
  journal={Nature},
  volume={549},
  number={7671},
  pages={195--202},
  year={2017},
  publisher={Nature Publishing Group},
  url = {https://doi.org/10.1038/nature23474}
}

@article{west2023towards,
  title={Towards quantum enhanced adversarial robustness in machine learning},
  author={West, Maxwell T and Tsang, Shu-Lok and Low, Jia S and Hill, Charles D and Leckie, Christopher and Hollenberg, Lloyd CL and Erfani, Sarah M and Usman, Muhammad},
  journal={Nature Machine Intelligence},
  volume={5},
  number={6},
  pages={581--589},
  year={2023},
  publisher={Nature Publishing Group UK London},
  DOI={https://doi.org/10.1038/s42256-023-00661-1}
}

@ARTICLE{8585034,
  author={Cao, Y. and Romero, J. and Aspuru-Guzik, A.},
  journal={IBM Journal of Research and Development}, 
  title={Potential of quantum computing for drug discovery}, 
  year={2018},
  volume={62},
  number={6},
  pages={6:1-6:20},
  doi={10.1147/JRD.2018.2888987}}

@article{robert2021resource,
  title={Resource-efficient quantum algorithm for protein folding},
  author={Robert, Anton and Barkoutsos, Panagiotis Kl and Woerner, Stefan and Tavernelli, Ivano},
  journal={npj Quantum Information},
  volume={7},
  number={1},
  pages={38},
  year={2021},
  publisher={Nature Publishing Group UK London},
  doi={https://doi.org/10.1038/s41534-021-00368-4}
}

@inproceedings{tang2019quantum,
author = {Tang, Ewin},
title = {A Quantum-Inspired Classical Algorithm for Recommendation Systems},
year = {2019},
isbn = {9781450367059},
publisher = {Association for Computing Machinery},
address = {New York, NY, USA},
url = {https://doi.org/10.1145/3313276.3316310},
doi = {10.1145/3313276.3316310},
booktitle = {Proceedings of the 51st Annual ACM SIGACT Symposium on Theory of Computing},
pages = {217–228},
numpages = {12},
location = {Phoenix, AZ, USA},
series = {STOC 2019}
}

@article{ladd2010quantum,
  title={Quantum computers},
  author={Ladd, Thaddeus D and Jelezko, Fedor and Laflamme, Raymond and Nakamura, Yasunobu and Monroe, Christopher and O’Brien, Jeremy Lloyd},
  journal={Nature},
  volume={464},
  number={7285},
  pages={45--53},
  year={2010},
  publisher={Nature Publishing Group},
  url = {https://doi.org/10.1038/nature08812}
}

@article{harrow2017quantum,
  title={Quantum computational supremacy},
  author={Harrow, Aram W and Montanaro, Ashley},
  journal={Nature},
  volume={549},
  number={7671},
  pages={203--209},
  year={2017},
  publisher={Nature Publishing Group},
  url = {https://doi.org/10.1038/nature23458}
}

@article{neill2018blueprint,
  title={A blueprint for demonstrating quantum supremacy with superconducting qubits},
  author={Neill, Charles and Roushan, Pedran and Kechedzhi, K and Boixo, Sergio and Isakov, Sergei V and Smelyanskiy, V and Megrant, A and Chiaro, B and Dunsworth, A and Arya, K and others},
  journal={Science},
  volume={360},
  number={6385},
  pages={195--199},
  year={2018},
  publisher={American Association for the Advancement of Science},
  url = {https://doi.org/10.1126/science.aao4309}
}

@article{yung2019quantum,
  title={Quantum supremacy: some fundamental concepts},
  author={Yung, Man-Hong},
  journal={National Science Review},
  volume={6},
  number={1},
  pages={22--23},
  year={2019},
  publisher={Oxford University Press},
  url = {https://doi.org/10.1093/nsr/nwy072}
}

@article{arute2019quantum,
  title={Quantum supremacy using a programmable superconducting processor},
  author={Arute, Frank and Arya, Kunal and Babbush, Ryan and Bacon, Dave and Bardin, Joseph C and Barends, Rami and Biswas, Rupak and Boixo, Sergio and Brandao, Fernando GSL and Buell, David A and others},
  journal={Nature},
  volume={574},
  number={7779},
  pages={505--510},
  year={2019},
  publisher={Nature Publishing Group},
  doi={https://doi.org/10.1038/s41586-019-1666-5}
}

@article{zhong2020quantum,
  title={Quantum computational advantage using photons},
  author={Zhong, Han-Sen and Wang, Hui and Deng, Yu-Hao and Chen, Ming-Cheng and Peng, Li-Chao and Luo, Yi-Han and Qin, Jian and Wu, Dian and Ding, Xing and Hu, Yi and others},
  journal={Science},
  volume={370},
  number={6523},
  pages={1460--1463},
  year={2020},
  publisher={American Association for the Advancement of Science},
  doi = {https://doi.org/10.1126/science.abe8770},
  url = {https://doi.org/10.1126/science.abe8770}
}

@article{PhysRevA.95.032307,
  title = {Coherence depletion in the Grover quantum search algorithm},
  author = {Shi, Hai-Long and Liu, Si-Yuan and Wang, Xiao-Hui and Yang, Wen-Li and Yang, Zhan-Ying and Fan, Heng},
  journal = {Phys. Rev. A},
  volume = {95},
  issue = {3},
  pages = {032307},
  numpages = {8},
  year = {2017},
  month = {Mar},
  publisher = {American Physical Society},
  doi = {10.1103/PhysRevA.95.032307},
  url = {https://link.aps.org/doi/10.1103/PhysRevA.95.032307}
}

@article{jozsa1998quantum,
  title={Quantum algorithms and the Fourier transform},
  author={Jozsa, Richard},
  journal={Phil. Trans. R. Soc. A.},
  volume={454},
  number={1969},
  pages={323--337},
  year={1998},
  publisher={The Royal Society},
  doi = {https://doi.org/10.1098/rspa.1998.0163},
  url = {https://doi.org/10.1098/rspa.1998.0163}
}

@article{ekert1998quantum,
  title={Quantum algorithms: entanglement--enhanced information processing},
  author={Ekert, Artur and Jozsa, Richard},
  journal={Phil. Trans. R. Soc. A.},
  volume={356},
  number={1743},
  pages={1769--1782},
  year={1998},
  publisher={The Royal Society},
  doi = {https://doi.org/10.1098/rsta.1998.0248},
  url = {https://doi.org/10.1098/rsta.1998.0248}
}

@article{PhysRevA.100.012349,
  title = {Operator coherence dynamics in Grover's quantum search algorithm},
  author = {Pan, Minghua and Qiu, Daowen},
  journal = {Phys. Rev. A},
  volume = {100},
  issue = {1},
  pages = {012349},
  numpages = {10},
  year = {2019},
  month = {Jul},
  publisher = {American Physical Society},
  doi = {10.1103/PhysRevA.100.012349},
  url = {https://link.aps.org/doi/10.1103/PhysRevA.100.012349}
}

@article{jozsa2003role,
  title={On the role of entanglement in quantum-computational speed-up},
  author={Jozsa, Richard and Linden, Noah},
  journal={Phil. Trans. R. Soc. A.},
  volume={459},
  number={2036},
  pages={2011--2032},
  year={2003},
  publisher={The Royal Society},
  url = {http://www.jstor.org/stable/3560059}
}

@article{ahnefeld2022coherence,
  title = {Coherence as a Resource for Shor's Algorithm},
  author = {Ahnefeld, Felix and Theurer, Thomas and Egloff, Dario and Matera, Juan Mauricio and Plenio, Martin B.},
  journal = {Phys. Rev. Lett.},
  volume = {129},
  issue = {12},
  pages = {120501},
  numpages = {7},
  year = {2022},
  month = {Sep},
  publisher = {American Physical Society},
  doi = {10.1103/PhysRevLett.129.120501},
  url = {https://link.aps.org/doi/10.1103/PhysRevLett.129.120501}
}

@article{pan2022complementarity,
  title={Complementarity between success probability and coherence in Grover search algorithm},
  author={Pan, Minghua and Situ, Haozhen and Zheng, Shenggen},
  journal={Europhysics Letters},
  volume={138},
  number={4},
  pages={48002},
  year={2022},
  publisher={IOP Publishing},
  doi = {10.1209/0295-5075/ac7165},
  url = {https://doi.org/10.1209/0295-5075/ac7165}
}

@article{PhysRevA.54.3824,
  title = {Mixed-state entanglement and quantum error correction},
  author = {Bennett, Charles H. and DiVincenzo, David P. and Smolin, John A. and Wootters, William K.},
  journal = {Phys. Rev. A},
  volume = {54},
  issue = {5},
  pages = {3824--3851},
  numpages = {0},
  year = {1996},
  month = {Nov},
  publisher = {American Physical Society},
  doi = {10.1103/PhysRevA.54.3824},
  url = {https://link.aps.org/doi/10.1103/PhysRevA.54.3824}
}

@article{PhysRevA.54.1838,
  title = {Information-theoretic aspects of inseparability of mixed states},
  author = {Horodecki, Ryszard and Horodecki, Michal/},
  journal = {Phys. Rev. A},
  volume = {54},
  issue = {3},
  pages = {1838--1843},
  numpages = {0},
  year = {1996},
  month = {Sep},
  publisher = {American Physical Society},
  doi = {10.1103/PhysRevA.54.1838},
  url = {https://link.aps.org/doi/10.1103/PhysRevA.54.1838}
}

@article{PhysRevA.60.1888,
  title = {General teleportation channel, singlet fraction, and quasidistillation},
  author = {Horodecki, Micha\l{} and Horodecki, Pawe\l{} and Horodecki, Ryszard},
  journal = {Phys. Rev. A},
  volume = {60},
  issue = {3},
  pages = {1888--1898},
  numpages = {0},
  year = {1999},
  month = {Sep},
  publisher = {American Physical Society},
  doi = {10.1103/PhysRevA.60.1888},
  url = {https://link.aps.org/doi/10.1103/PhysRevA.60.1888}
}

@article{PhysRevA.62.012311,
  title = {Local environment can enhance fidelity of quantum teleportation},
  author = {Badzia\ifmmode \mbox{\c{}}\else \c{}\fi{}g, Piotr and Horodecki, Micha\l{} and Horodecki, Pawe\l{} and Horodecki, Ryszard},
  journal = {Phys. Rev. A},
  volume = {62},
  issue = {1},
  pages = {012311},
  numpages = {7},
  year = {2000},
  month = {Jun},
  publisher = {American Physical Society},
  doi = {10.1103/PhysRevA.62.012311},
  url = {https://link.aps.org/doi/10.1103/PhysRevA.62.012311}
}

@article{PhysRevA.66.012301,
  title = {Optimal teleportation based on bell measurements},
  author = {Albeverio, Sergio and Fei, Shao-Ming and Yang, Wen-Li},
  journal = {Phys. Rev. A},
  volume = {66},
  issue = {1},
  pages = {012301},
  numpages = {4},
  year = {2002},
  month = {Jul},
  publisher = {American Physical Society},
  doi = {10.1103/PhysRevA.66.012301},
  url = {https://link.aps.org/doi/10.1103/PhysRevA.66.012301}
}

@article{PhysRevA.91.012310,
  title = {Maximally entangled states and fully entangled fraction},
  author = {Zhao, Ming-Jing},
  journal = {Phys. Rev. A},
  volume = {91},
  issue = {1},
  pages = {012310},
  numpages = {5},
  year = {2015},
  month = {Jan},
  publisher = {American Physical Society},
  doi = {10.1103/PhysRevA.91.012310},
  url = {https://link.aps.org/doi/10.1103/PhysRevA.91.012310}
}

@article{PhysRevA.64.022307,
  title = {Grover algorithm with zero theoretical failure rate},
  author = {Long, G. L.},
  journal = {Phys. Rev. A},
  volume = {64},
  issue = {2},
  pages = {022307},
  numpages = {4},
  year = {2001},
  month = {Jul},
  publisher = {American Physical Society},
  doi = {10.1103/PhysRevA.64.022307},
  url = {https://link.aps.org/doi/10.1103/PhysRevA.64.022307}
}

@article{PhysRevA.68.022326,
  title = {Analysis of Grover's quantum search algorithm as a dynamical system},
  author = {Biham, Ofer and Shapira, Daniel and Shimoni, Yishai},
  journal = {Phys. Rev. A},
  volume = {68},
  issue = {2},
  pages = {022326},
  numpages = {8},
  year = {2003},
  month = {Aug},
  publisher = {American Physical Society},
  doi = {10.1103/PhysRevA.68.022326},
  url = {https://link.aps.org/doi/10.1103/PhysRevA.68.022326}
}

@article{PhysRevA.71.042320,
  title = {Algebraic analysis of quantum search with pure and mixed states},
  author = {Shapira, Daniel and Shimoni, Yishai and Biham, Ofer},
  journal = {Phys. Rev. A},
  volume = {71},
  issue = {4},
  pages = {042320},
  numpages = {8},
  year = {2005},
  month = {Apr},
  publisher = {American Physical Society},
  doi = {10.1103/PhysRevA.71.042320},
  url = {https://link.aps.org/doi/10.1103/PhysRevA.71.042320}
}

@article{li2023deterministic,
  title={Deterministic quantum search with adjustable parameters: Implementations and applications},
  author={Li, Guanzhong and Li, Lvzhou},
  journal={Information and Computation},
  volume={292},
  pages={105042},
  year={2023},
  publisher={Elsevier},
doi={10.1016/j.ic.2023.105042},
url={https://doi.org/10.1016/j.ic.2023.105042}

}

@misc{zhou2024distributedexactgeneralizedgrovers,
  title={Distributed Exact Generalized Grover's Algorithm}, 
  author={Xu Zhou and Xusheng Xu and Shenggen Zheng and Le Luo},
  year={2024},
  eprint={2405.06963},
  archivePrefix={arXiv},
  primaryClass={quant-ph},
  url={https://arxiv.org/abs/2405.06963}, 
}

@article{PhysRevA.100.032324,
  title = {Quantum coherence fraction},
  author = {Yao, Yao and Li, Dong and Sun, C. P.},
  journal = {Phys. Rev. A},
  volume = {100},
  issue = {3},
  pages = {032324},
  numpages = {7},
  year = {2019},
  month = {Sep},
  publisher = {American Physical Society},
  doi = {10.1103/PhysRevA.100.032324},
  url = {https://link.aps.org/doi/10.1103/PhysRevA.100.032324}
}

@article{karmakar2019coherence,
  title={Coherence fraction},
  author={Karmakar, Sumana and Sen, Ajoy and Chattopadhyay, Indrani and Bhar, Amit and Sarkar, Debasis},
  journal={Quantum Information Processing},
  volume={18},
  number={9},
  pages={275},
  year={2019},
  publisher={Springer},
doi={https://doi.org/10.1007/s11128-019-2391-6}
}

@article{PhysRevLett.80.4329,
  title = {Quantum Computers Can Search Rapidly by Using Almost Any Transformation},
  author = {Grover, Lov K.},
  journal = {Phys. Rev. Lett.},
  volume = {80},
  issue = {19},
  pages = {4329--4332},
  numpages = {0},
  year = {1998},
  month = {May},
  publisher = {American Physical Society},
  doi = {10.1103/PhysRevLett.80.4329},
  url = {https://link.aps.org/doi/10.1103/PhysRevLett.80.4329}
}

@article{PhysRevA.65.062312,
  title = {Entanglement monotone derived from Grover's algorithm},
  author = {Biham, Ofer and Nielsen, Michael A. and Osborne, Tobias J.},
  journal = {Phys. Rev. A},
  volume = {65},
  issue = {6},
  pages = {062312},
  numpages = {7},
  year = {2002},
  month = {Jun},
  publisher = {American Physical Society},
  doi = {10.1103/PhysRevA.65.062312},
  url = {https://link.aps.org/doi/10.1103/PhysRevA.65.062312}
}

@article{PhysRevA.93.012111,
  title = {Coherence as a resource in decision problems: The Deutsch-Jozsa algorithm and a variation},
  author = {Hillery, Mark},
  journal = {Phys. Rev. A},
  volume = {93},
  issue = {1},
  pages = {012111},
  numpages = {6},
  year = {2016},
  month = {Jan},
  publisher = {American Physical Society},
  doi = {10.1103/PhysRevA.93.012111},
  url = {https://link.aps.org/doi/10.1103/PhysRevA.93.012111}
}

@misc{anand2016coherenceentanglementmonogamydiscrete,
      title={Coherence and Entanglement Monogamy in the Discrete Analogue of Analog Grover Search}, 
      author={Namit Anand and Arun Kumar Pati},
      year={2016},
      eprint={1611.04542},
      archivePrefix={arXiv},
      primaryClass={quant-ph},
      url={https://arxiv.org/abs/1611.04542}, 
}

@article{liu2019coherence,
  title={Coherence depletion in quantum algorithms},
  author={Liu, Ye-Chao and Shang, Jiangwei and Zhang, Xiangdong},
  journal={Entropy},
  volume={21},
  number={3},
  pages={260},
  year={2019},
  publisher={MDPI},
  url = {https://doi.org/10.3390/e21030260}
}

@article{RevModPhys.81.865,
  title = {Quantum entanglement},
  author = {Horodecki, Ryszard and Horodecki, Pawe\l{} and Horodecki, Micha\l{} and Horodecki, Karol},
  journal = {Rev. Mod. Phys.},
  volume = {81},
  issue = {2},
  pages = {865--942},
  numpages = {0},
  year = {2009},
  month = {Jun},
  publisher = {American Physical Society},
  doi = {10.1103/RevModPhys.81.865},
  url = {https://link.aps.org/doi/10.1103/RevModPhys.81.865}
}

@article{guhne2009entanglement,
  title={Entanglement detection},
  author={G{\"u}hne, Otfried and T{\'o}th, G{\'e}za},
  journal={Physics Reports},
  volume={474},
  number={1-6},
  pages={1--75},
  year={2009},
  publisher={Elsevier},
  doi = {10.1016/j.physrep.2009.02.004},
  url = {https://doi.org/10.1016/j.physrep.2009.02.004}
}

@article{gour2008resource,
  title={The resource theory of quantum reference frames: manipulations and monotones},
  author={Gour, Gilad and Spekkens, Robert W},
  journal={New Journal of Physics},
  volume={10},
  number={3},
  pages={033023},
  year={2008},
  publisher={IOP Publishing},
  url = {https://doi.org/10.1088/1367-2630/10/3/033023}
}

@article{marvian2014extending,
  title={Extending Noether’s theorem by quantifying the asymmetry of quantum states},
  author={Marvian, Iman and Spekkens, Robert W},
  journal={Nature communications},
  volume={5},
  number={1},
  pages={1--8},
  year={2014},
  publisher={Nature Publishing Group},
  url = {https://doi.org/10.1038/ncomms4821}
}

@article{PhysRevLett.113.140401,
  title = {Quantifying Coherence},
  author = {Baumgratz, T. and Cramer, M. and Plenio, M. B.},
  journal = {Phys. Rev. Lett.},
  volume = {113},
  issue = {14},
  pages = {140401},
  numpages = {5},
  year = {2014},
  month = {Sep},
  publisher = {American Physical Society},
  doi = {10.1103/PhysRevLett.113.140401},
  url = {https://link.aps.org/doi/10.1103/PhysRevLett.113.140401}
}

@article{henderson2001classical,
  title={Classical, quantum and total correlations},
  author={Henderson, Leah and Vedral, Vlatko},
  journal={Journal of physics A: mathematical and general},
  volume={34},
  number={35},
  pages={6899},
  year={2001},
  publisher={IOP Publishing},
doi={10.1088/0305-4470/34/35/315}
}

@article{PhysRevLett.88.017901,
  title = {Quantum Discord: A Measure of the Quantumness of Correlations},
  author = {Ollivier, Harold and Zurek, Wojciech H.},
  journal = {Phys. Rev. Lett.},
  volume = {88},
  issue = {1},
  pages = {017901},
  numpages = {4},
  year = {2001},
  month = {Dec},
  publisher = {American Physical Society},
  doi = {10.1103/PhysRevLett.88.017901},
  url = {https://link.aps.org/doi/10.1103/PhysRevLett.88.017901}
}

@article{PhysRevLett.100.050502,
  title = {Quantum Discord and the Power of One Qubit},
  author = {Datta, Animesh and Shaji, Anil and Caves, Carlton M.},
  journal = {Phys. Rev. Lett.},
  volume = {100},
  issue = {5},
  pages = {050502},
  numpages = {4},
  year = {2008},
  month = {Feb},
  publisher = {American Physical Society},
  doi = {10.1103/PhysRevLett.100.050502},
  url = {https://link.aps.org/doi/10.1103/PhysRevLett.100.050502}
}

@misc{durr1996quantum,
  title={A Quantum Algorithm for Finding the Minimum}, 
  author={Christoph Durr and Peter Hoyer},
  year={1999},
  eprint={quant-ph/9607014},
  archivePrefix={arXiv},
  primaryClass={quant-ph},
  url={https://arxiv.org/abs/quant-ph/9607014}, 
}

@book{wittek2014quantum,
  title={Quantum machine learning: what quantum computing means to data mining},
  author={Wittek, Peter},
  year={2014},
  publisher={Academic Press},
  url = {https://doi.org/10.1016/C2013-0-19170-2}
}

@article{PhysRevLett.113.130503,
  title = {Quantum Support Vector Machine for Big Data Classification},
  author = {Rebentrost, Patrick and Mohseni, Masoud and Lloyd, Seth},
  journal = {Phys. Rev. Lett.},
  volume = {113},
  issue = {13},
  pages = {130503},
  numpages = {5},
  year = {2014},
  month = {Sep},
  publisher = {American Physical Society},
  doi = {10.1103/PhysRevLett.113.130503},
  url = {https://link.aps.org/doi/10.1103/PhysRevLett.113.130503}
}

@article{schuld2015introduction,
  title={An introduction to quantum machine learning},
  author={Schuld, Maria and Sinayskiy, Ilya and Petruccione, Francesco},
  journal={Contemporary Physics},
  volume={56},
  number={2},
  pages={172--185},
  year={2015},
  publisher={Taylor \& Francis},
  url = {https://doi.org/10.1080/00107514.2014.964942}
}

@article{boyer1998tight,
  title={Tight bounds on quantum searching},
  author={Boyer, Michel and Brassard, Gilles and H{\o}yer, Peter and Tapp, Alain},
  journal={Fortschritte der Physik: Progress of Physics},
  volume={46},
  number={4-5},
  pages={493--505},
  year={1998},
  publisher={Wiley Online Library},
  doi = {https://doi.org/10.1002/(SICI)1521-3978(199806)46:4/5<493::AID-PROP493>3.0.CO;2-P},
  url = {https://onlinelibrary.wiley.com/doi/abs/10.1002/%28SICI%291521-3978%28199806%2946%3A4/5%3C493%3A%3AAID-PROP493%3E3.0.CO%3B2-P
  },
}

@article{tang2021quantum,
  title = {Quantum Principal Component Analysis Only Achieves an Exponential Speedup Because of Its State Preparation Assumptions},
  author = {Tang, Ewin},
  journal = {Phys. Rev. Lett.},
  volume = {127},
  issue = {6},
  pages = {060503},
  numpages = {6},
  year = {2021},
  month = {Aug},
  publisher = {American Physical Society},
  doi = {10.1103/PhysRevLett.127.060503},
  url = {https://link.aps.org/doi/10.1103/PhysRevLett.127.060503}
}

@article{shao2024simulating,
  title = {Simulating Noisy Variational Quantum Algorithms: A Polynomial Approach},
  author = {Shao, Yuguo and Wei, Fuchuan and Cheng, Song and Liu, Zhengwei},
  journal = {Phys. Rev. Lett.},
  volume = {133},
  issue = {12},
  pages = {120603},
  numpages = {7},
  year = {2024},
  month = {Sep},
  publisher = {American Physical Society},
  doi = {10.1103/PhysRevLett.133.120603},
  url = {https://link.aps.org/doi/10.1103/PhysRevLett.133.120603}
}

@article{bennett1997strengths,
  title={Strengths and weaknesses of quantum computing},
  author={Bennett, Charles H and Bernstein, Ethan and Brassard, Gilles and Vazirani, Umesh},
  journal={SIAM journal on Computing},
  volume={26},
  number={5},
  pages={1510--1523},
  year={1997},
  publisher={SIAM},
doi={10.1137/S0097539796300933},
url ={https://doi.org/10.1137/S0097539796300933}
}

@article{PhysRevA.60.2746,
  title = {Grover's quantum searching algorithm is optimal},
  author = {Zalka, Christof},
  journal = {Phys. Rev. A},
  volume = {60},
  issue = {4},
  pages = {2746--2751},
  numpages = {0},
  year = {1999},
  month = {Oct},
  publisher = {American Physical Society},
  doi = {10.1103/PhysRevA.60.2746},
  url = {https://link.aps.org/doi/10.1103/PhysRevA.60.2746}
}

@article{Gilliam2021groveradaptive,
  doi = {10.22331/q-2021-04-08-428},
  url = {https://doi.org/10.22331/q-2021-04-08-428},
  title = {Grover {A}daptive {S}earch for {C}onstrained {P}olynomial {B}inary {O}ptimization},
  author = {Gilliam, Austin and Woerner, Stefan and Gonciulea, Constantin},
  journal = {{Quantum}},
  issn = {2521-327X},
  publisher = {{Verein zur F{\"{o}}rderung des Open Access Publizierens in den Quantenwissenschaften}},
  volume = {5},
  pages = {428},
  month = apr,
  year = {2021}
}

@article{ohno2024grover,
  title={Grover’s search with learning oracle for constrained binary optimization problems},
  author={Ohno, Hiroshi},
  journal={Quantum Machine Intelligence},
  volume={6},
  number={1},
  pages={12},
  year={2024},
  publisher={Springer},
  doi={https://doi.org/10.1007/s42484-024-00148-1}
}

@article{PhysRevLett.118.020403,
  title = {Directly Measuring the Degree of Quantum Coherence using Interference Fringes},
  author = {Wang, Yi-Tao and Tang, Jian-Shun and Wei, Zhi-Yuan and Yu, Shang and Ke, Zhi-Jin and Xu, Xiao-Ye and Li, Chuan-Feng and Guo, Guang-Can},
  journal = {Phys. Rev. Lett.},
  volume = {118},
  issue = {2},
  pages = {020403},
  numpages = {5},
  year = {2017},
  month = {Jan},
  publisher = {American Physical Society},
  doi = {10.1103/PhysRevLett.118.020403},
  url = {https://link.aps.org/doi/10.1103/PhysRevLett.118.020403}
}

@article{RevModPhys.89.041003,
  title = {Colloquium: Quantum coherence as a resource},
  author = {Streltsov, Alexander and Adesso, Gerardo and Plenio, Martin B.},
  journal = {Rev. Mod. Phys.},
  volume = {89},
  issue = {4},
  pages = {041003},
  numpages = {34},
  year = {2017},
  month = {Oct},
  publisher = {American Physical Society},
  doi = {10.1103/RevModPhys.89.041003},
  url = {https://link.aps.org/doi/10.1103/RevModPhys.89.041003}
}

@article{PhysRevLett.120.230504,
  title = {Experimental Demonstration of Observability and Operability of Robustness of Coherence},
  author = {Zheng, Wenqiang and Ma, Zhihao and Wang, Hengyan and Fei, Shao-Ming and Peng, Xinhua},
  journal = {Phys. Rev. Lett.},
  volume = {120},
  issue = {23},
  pages = {230504},
  numpages = {5},
  year = {2018},
  month = {Jun},
  publisher = {American Physical Society},
  doi = {10.1103/PhysRevLett.120.230504},
  url = {https://link.aps.org/doi/10.1103/PhysRevLett.120.230504}
}

@article{hu2018quantum,
  title={Quantum coherence and geometric quantum discord},
  author={Hu, Ming-Liang and Hu, Xueyuan and Wang, Jieci and Peng, Yi and Zhang, Yu-Ran and Fan, Heng},
  journal={Physics Reports},
  volume={762},
  pages={1--100},
  year={2018},
  publisher={Elsevier},
doi={10.1016/j.physrep.2018.07.004}
}

@article{PhysRevLett.125.060404,
  title = {Experimental Quantification of Coherence of a Tunable Quantum Detector},
  author = {Xu, Huichao and Xu, Feixiang and Theurer, Thomas and Egloff, Dario and Liu, Zi-Wen and Yu, Nengkun and Plenio, Martin B. and Zhang, Lijian},
  journal = {Phys. Rev. Lett.},
  volume = {125},
  issue = {6},
  pages = {060404},
  numpages = {7},
  year = {2020},
  month = {Aug},
  publisher = {American Physical Society},
  doi = {10.1103/PhysRevLett.125.060404},
  url = {https://link.aps.org/doi/10.1103/PhysRevLett.125.060404}
}

@article{wu2021experimental,
  title={Experimental progress on quantum coherence: detection, quantification, and manipulation},
  author={Wu, Kang-Da and Streltsov, Alexander and Regula, Bartosz and Xiang, Guo-Yong and Li, Chuan-Feng and Guo, Guang-Can},
  journal={Advanced Quantum Technologies},
  volume={4},
  number={9},
  pages={2100040},
  year={2021},
  publisher={Wiley Online Library},
  doi={10.1002/qute.202100040}
}

@article{yang2023device,
  title={Device-independent verification of quantum coherence without quantum control},
  author={Yang, Yan-Han and Yang, Xue and Zheng, Xing-Zhou and Luo, Ming-Xing},
  journal={Cell Reports Physical Science},
  volume={4},
  number={12},
  pages={101725},
  year={2023},
  doi={10.1016/j.xcrp.2023.101725},
  publisher={Elsevier}
}

\end{document}